\documentclass[superscriptaddress,twocolumn,prl]{revtex4-1}
\usepackage[latin9]{inputenc}
\usepackage{amsmath}
\usepackage{upgreek}
\usepackage{amstext}
\usepackage{graphicx}  
\usepackage{changes}
\usepackage{array}

\begin{document}
\title{Test of the Equivalence Principle for Superconductors}
\author{M.P. Ross}
\affiliation{Center for Experimental Nuclear Physics and Astrophysics, University of Washington, Seattle, Washington 98195, USA}
\author{S.M. Fleischer}
\affiliation{Department of Physics and Astronomy,\\ Western Washington University, Bellingham, Washington 98225, USA}
\author{ I.A. Paulson}
\affiliation{Department of Physics and Astronomy,\\ Carleton College, Northfield, Minnesota 98133, USA}
\author{ P. Lamb}
\affiliation{Department of Physics and Astronomy,\\ Western Washington University, Bellingham, Washington 98225, USA}
\author{ B.M. Iritani}
\affiliation{Department of Physics,\\ Columbia University, New York, NY 10027, USA}
\author{  E.G. Adelberger}
\author{ C.A. Hagedorn}
\author{ K. Venkateswara}
\author{ C. Gettings}
\author{ E.A. Shaw}
\author{ S.K.Apple}
\author{ J.H. Gundlach}
\affiliation{Center for Experimental Nuclear Physics and Astrophysics, University of Washington, Seattle, Washington 98195, USA}

\begin{abstract}
We searched for violations of the weak equivalence principle using a cryogenic torsion balance with a pendulum comprised of superconducting niobium and copper. We constrain the E\"otv\"os parameter with 95\%-confidence to $\eta_{\text{Nb*-Cu}}~\leq~2.0\times10^{-9}$ and $\eta_{\text{CP-ee}}\leq9.2\times10^{-4}$ for superconducting niobium and Cooper pairs, respectively.
\end{abstract}

\maketitle

The weak equivalence principle (WEP) underpins our geometric description of gravity (\textit{i.e.} general relativity). Possible violations of the WEP are traditionally parameterized by the E\"otv\"os parameter \cite{Eotvos}: 
\begin{equation}
\eta=\frac{2(a_1-a_2)}{a_1+a_2}
\end{equation}
where $a_{1,2}$ are the gravitational accelerations for two test bodies made from different materials. The WEP states that this parameter vanishes for all choices of material. This has been precisely tested for a variety of materials. \cite{NewWash, LLR, MICROSCOPE}.

It has been argued \cite{superTheory,GCasimir, GWReflect} that the non-local quantum-mechanical properties of Cooper pairs should lead to violations of the WEP, with sometimes remarkable effects. For example, it has been suggested that superconductors strongly reflect gravitational waves \cite{GCasimir, GWReflect}, which would imply a major violation of the WEP.

To our knowledge, only two previous experiments \cite{Jain, Tajmar} have probed the gravitational properties of superconductors. In particular, \citet{Tajmar} compared the change in the inertial and gravitational masses of niobium in the normal and superconducting states and obtained $\eta_{\text{Nb*-Cu}} < 4.0\times 10^{-7}.$

Here, we describe a sensitive test of the WEP comparing superconducting niobium with normally-conducting copper. The experiment was made possible by the development of a unique cryogenic, torsion balance apparatus.  

\begin{figure}[!h]
\centering \includegraphics[width=0.5\textwidth]{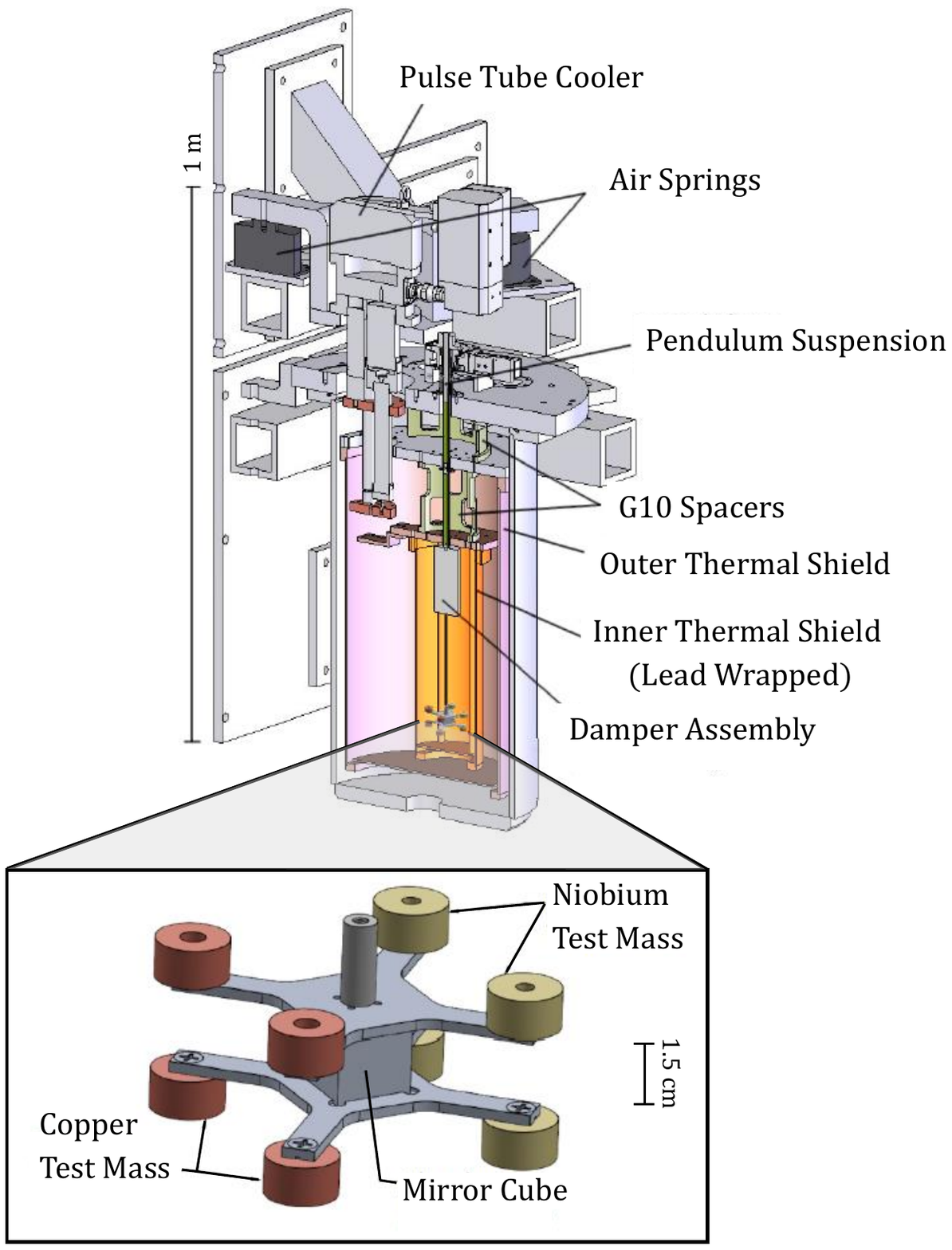}
\caption{Schematic cross-sectional view of the torsion balance apparatus. Some parts have been omitted, including the magnetic shields and the autocollimator. Inset shows a detail of the torsion pendulum. Adapted from \citet{RSI}, with the permission of AIP Publishing.}
\label{apparatus} 
\end{figure}

\textit{Apparatus} - Very few torsion balances have operated at cryogenic temperatures \cite{otherCryo}. Our cryogenic balance (described in detail in \citet{RSI}) is shown in Figure~\ref{apparatus}. 
The balance consisted of a torsion pendulum with a niobium-copper composition dipole suspended from a 30-{\textmu}m thick, 16.5-cm long tungsten fiber which was attached to a passively isolated ``swing" damper. The pendulum was housed in a vacuum chamber with a 0.4 mPa helium environment and surrounded by two layers of thermal shielding. The innermost layer of shielding and the suspension point of the fiber were cooled to $\sim$~5 K by a two-stage helium pulse-tube cooler. An outer shield was cooled to $\sim$~50 K. The exterior of the inner shield was wrapped in a cylindrical layer of lead to provide superconducting magnetic shielding when at cryogenic temperatures. Additionally, a traditional magnetic shield (Co-NETIC AA) surrounded the outer thermal shield.

The pendulum had eight test bodies each with a mass of 9.5 g, four made of OFHC copper and four of niobium (99.9\% purity), in a dipole arrangement. The parameters that describe the torsion balance are displayed in Table \ref{appTable}. When the pendulum was cooled below the critical temperature of niobium, 9~K~\cite{Nb}, the test bodies formed a superconductor-normal-conductor dipole. The superconductivity of the niobium was verified by demonstrating the Meissner effect with an auxiliary coil placed near the pendulum. The angular position of the pendulum was measured by a multi-slit autocollimator \cite{MSA}. 

\begin{table}

\begin{tabular}{ | m{0.2\textwidth} | m{0.05\textwidth}|| m{0.2\textwidth}|}
 \hline
 \multicolumn{3}{|c|}{Apparatus} \\
 \hline
 Parameter &  & Value \\
 \hline
 Spring Constant & \centering{$\kappa$}  &\ $  8.7 \times 10^{-8}$ N m/rad\\
 Mass of one test body set & \centering{$m$}  &\ $  38 $ g\\
 Moment of inertia & \centering{$I$}  &\ $  8.6 \times 10^{-5}$ kg $\text{m}^2$\\
 Resonant frequency & \centering{$\omega_0$}  &\ $  2\pi \times5.1$ mHz\\
 Lever-arm & \centering{$r$}  &\ $  2.3 $ cm\\
 \hline
\end{tabular}
 \caption{Table 1: Relevant parameters of the torsion balance. The lever-arm $r$ is the radial distance from the pendulum's vertical axis to the center of mass of each set of test bodies.}\label{appTable}

\end{table}

Environmental sensors were placed throughout the instrument. These included an orthogonal set of tiltmeters attached to the top flange of the vacuum chamber, a pair of external magnetometers, and an array of temperature sensors placed on the top flange, first and second stage of the cold head, and inner and outer thermal shields.

\begin{figure}[!h]
\centering \includegraphics[width=0.49\textwidth]{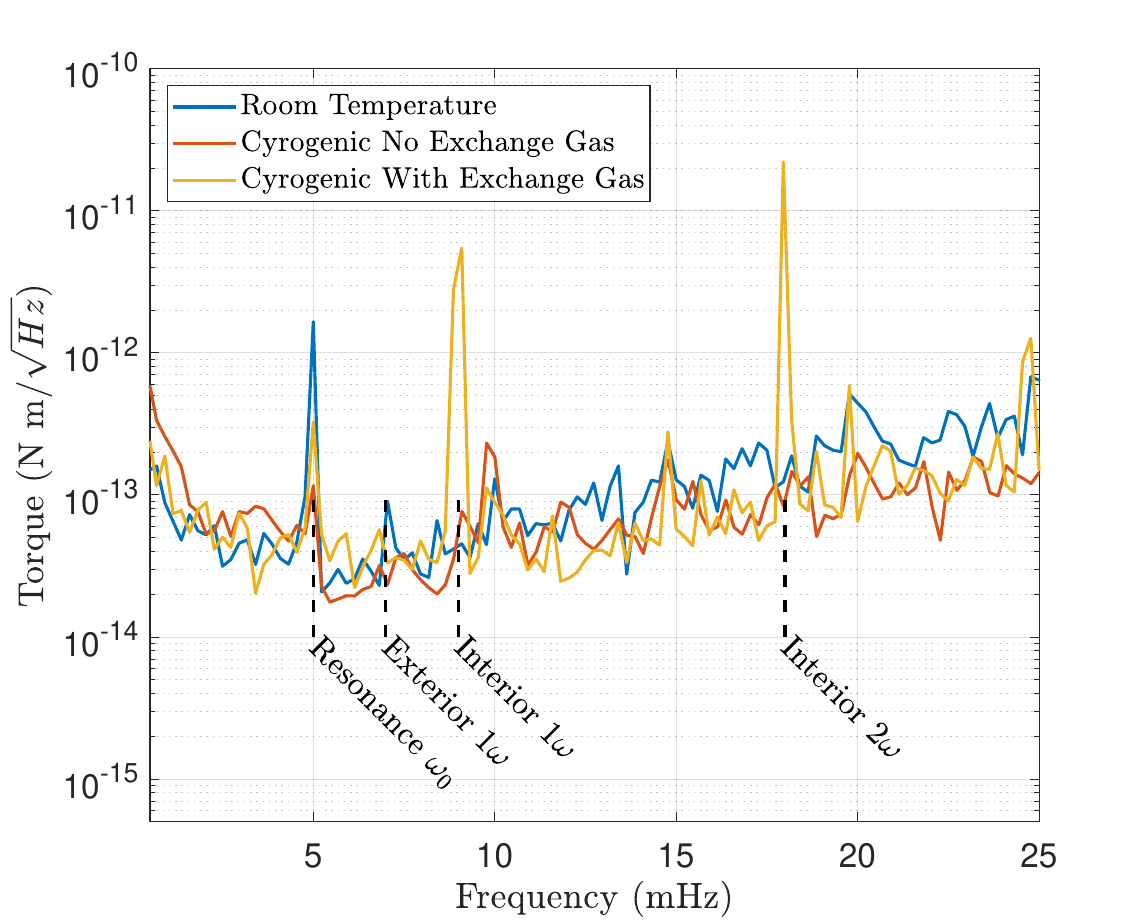}
\caption{Verification of the superconducting state of the niobium test bodies and magnetic shield via sinusoidal magnetic field injection from inside the chamber (9 mHz) and outside the chamber (7 mHz).}\label{super} 
\end{figure}

\textit{Superconductivity of Niobium Test Bodies}\label{SuperSup} - A direct measurement of the pendulum temperature was not practical without spoiling the performance of the torsion balance. Although the inner shield is at $\sim$ 5 K, the equilibrium temperature of the pendulum at low pressures ($<$0.1 mPa) is above the critical temperature of niobium. This is due to the heat loads from the autocollimator light and residual thermal radiation from room temperature surfaces visible to the pendulum through the optical access window.

We achieved superconductivity by injecting helium exchange gas into the chamber to a final pressure of 0.4 mPa. This pressure allowed sufficient thermal coupling between the pendulum and the inner shield to bring the pendulum below the critical temperature of niobium.

We verified that the test bodies were superconducting by observing the pendulum's response to an injected sinusoidal magnetic field with coils placed inside and outside the chamber. The torque acting on the pendulum due to an external magnetic field, $\vec{B}$, has three terms: one due to the intrinsic magnetic moment of the pendulum, $\vec{m}_0$, a second due to an effective magnetic moment induced by pinned flux, $\vec{m}_\text{flux}$, and a third due to the expulsion of the magnetic field. These are described by:
\begin{equation}
    \vec{\tau}=(\vec{m}_0+\vec{m}_\text{flux})\times \vec{B}-\vec{r}\times\bigg(\int \frac{B^2}{2 \mu_0} d\vec{A}\bigg).
\end{equation}

An applied sinusoidal magnetic field, $\vec{B}=\vec{B}_0 \sin \omega t$, produces three distinct $\omega$-dependent torques:

\begin{align}
    \vec{\tau}=&\vec{\tau}_0 \sin \omega t +\vec{\tau}_\text{flux} \sin \omega t \\
    &+\vec{\tau}_\text{Meissner}(\cos 2\omega t)\nonumber
\end{align}
where $\vec{\tau}_0,\ \vec{\tau}_\text{flux},\ \vec{\tau}_\text{Meissner}$ are the torque amplitudes due to the magnetic moment of the pendulum, the pinned flux, and the Meissner effect, respectively.

A response at the drive frequency, $\omega$, is expected when the test bodies are not superconducting as both the pinned flux and Meissner terms do not contribute. However, when the test bodies are superconducting, an increase in the response at the drive frequency and a large response at twice the drive frequency, $2\omega$, are expected.

Figure \ref{super} shows the measured torque spectra at room temperature as well as at low temperature, with and without exchange gas, while magnetic fields were injected via exterior and interior coils. With the addition of the exchange gas, the pendulum has a large $2\omega$ response to the interior coil due to the Meissner effect and an increased $1\omega$ response due to pinned flux, confirming that the niobium test bodies are in a superconducting state. Additionally, the response to the exterior field vanishes due to the lead shield becoming superconducting during cryogenic operations.

\begin{widetext}

\begin{figure}[!h]
\centering \includegraphics[width=\textwidth]{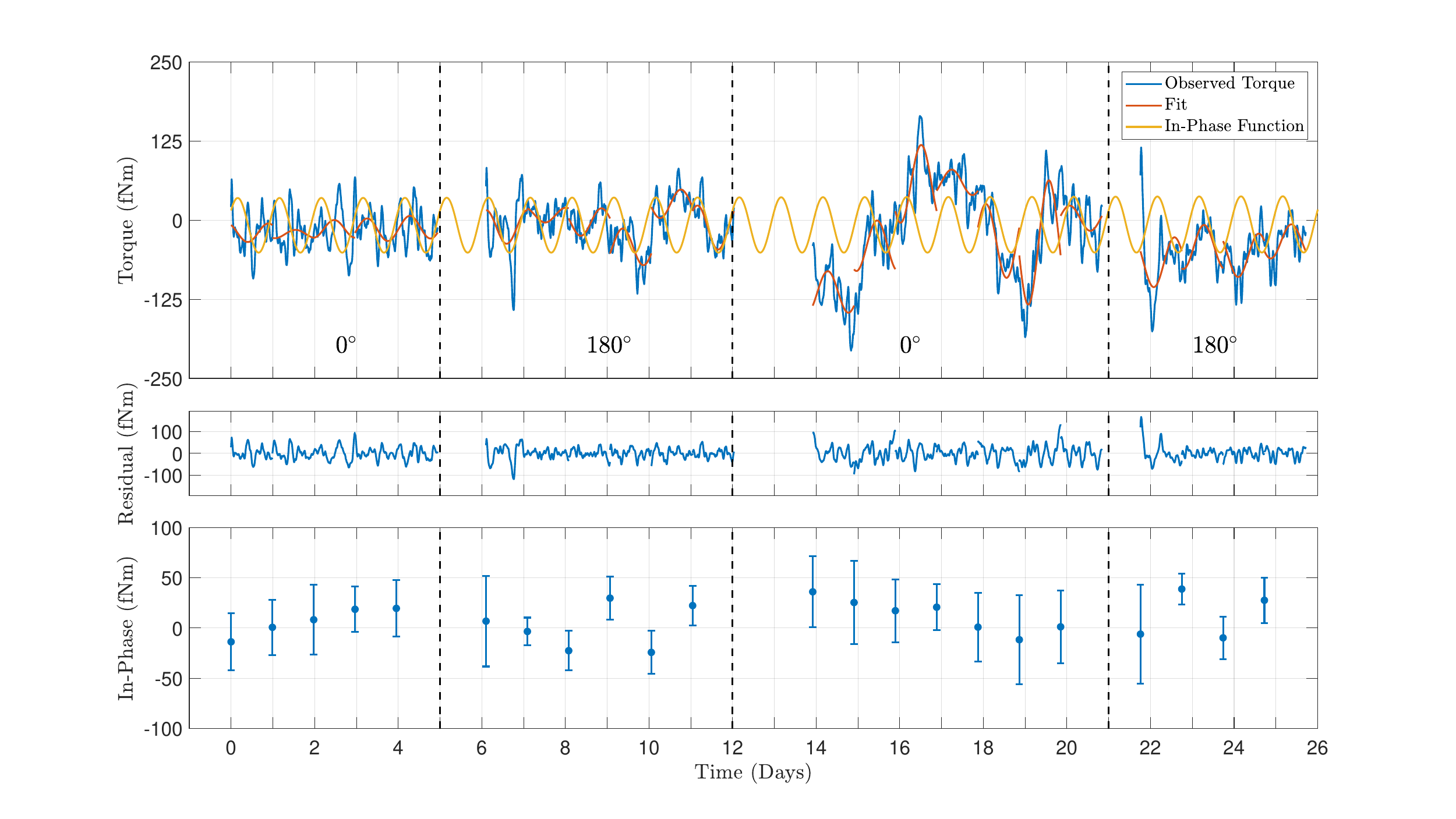}
\caption{Time series of the observed temperature-corrected torque, the daily fits, and the in-phase function for comparison (scaled assuming $\eta_{\text{Nb}}=10^{-8}$), as well as the residual when the fits are subtracted from the data. Gaps in the data are due to the thermalization and damping required after pendulum flips as well as a failure of the lab's air handling system on day thirteen. The bottom pane shows the time series of the in-phase torque fit amplitudes and 1$\sigma$ error bars. Each cut is one day long. The vertical dashed lines indicate pendulum flips.}\label{timeSeries} 
\end{figure}

\end{widetext}

\textit{Analysis} - The experiment was operated from March 26, 2024 to April 22, 2024 with occasional disruptions, yielding 22 days of usable data. We searched for a violation of the equivalence principle towards the sun that would appear as a daily modulated torque. Our analysis began by filtering the observed torque with a second-order low-pass Butterworth filter with a corner frequency of 0.1~mHz. This filter removes high-frequency noise caused by the angular readout as well as the free torsional oscillation. 

The raw torque showed a clear correlation with the temperature of the inner shield. We interpret this coupling as thermally induced torques of the pendulum's suspension point. This correlation was removed by subtracting the inner shield temperature scaled by a constant coupling factor. This coupling factor was determined to be $4.6 \times 10^{-11}$ N m/K by correlating the raw torque and inner shield temperature across the entire data run. Additionally, coupling to the tilt of the apparatus was subtracted from the measured torque. This subtraction did not significantly change the observed torque. The tilt coupling was determined by deliberately tilting the chamber to be $-5.5 \times 10^{-8}$~N~m/rad and $3.5 \times 10^{-8}$~N~m/rad respectively for the two horizontal directions. An upper limit of $10^{-11}$~N~m/Gauss was placed on coupling to external magnetic fields during the magnetic injections. This indicates a negligible influence ($<0.1$ fN m) of external magnetic fields on the observed torque. None of the environmental couplings changed appreciably throughout the data run.

We periodically rotated the pendulum by 180$^\circ$, effectively exchanging the copper and niobium test bodies, to minimize the environmental effects that could mimic an EP violation. An EP-violating torque would have the opposite sign when the pendulum is flipped whereas many environmental signals (daily tilts, etc.) would not change. This allows EP-violation-mimicking environmental effects to be subtracted from the results. The pendulum requires damping and a thermalization period after each rotation. As such, the data was recorded in roughly week-long sets. A third-order polynomial was fit and subtracted from each set to account for long term drifts.

A series of day-long cuts of the corrected torque were extracted and analyzed independently. Each cut was linear least-squares fitted to the sum of two functions of the sun's location and an offset term: 
\begin{align}
\tau (t)= &\tau_{in}\ a_{in}(t)+ \tau_{out}\ a_{out}(t)+ C_{offset}
\end{align}
where $\tau$ corresponds to the observed torque, $a_{in}$ is the ``in-phase" horizontal projection of the sun's location perpendicular to the pendulum's composition dipole, $a_{out}$ is the ``out-of-phase" orthogonally-oriented horizontal projection, $t$ is time, $\tau_{in}$ and $\tau_{out}$ are the respective torque fit coefficients, and $C_{offset}$ is the fit offset. 

The sun's location was obtained using the {\tt Astropy} \cite{astropy1, astropy2} libraries. Figure \ref{timeSeries} shows the observed and fit torque along with the daily modulated in-phase function and the residual of the fits. An EP violating signal would appear as a non-zero in-phase fit coefficient. 

Figure \ref{timeSeries} shows the collection of extracted in-phase coefficients. We take the standard deviation of the residual torque for each cut as a conservative estimate of the uncertainty of the corresponding torque fit coefficient. The collection of in-phase coefficients were averaged with inverse-variance weighting to yield a measurement of the in-phase torque. We find no evidence of a EP-violation with a measured in-phase torque amplitude of:
\begin{equation}
    \tau_{in}=-1.3\pm5.2\text{ fN m}
\end{equation}

We also measured no significant out-of-phase component, $\tau_{out}=-2.1\pm5.2\text{ fN m}$. We extract limits on the E\"otv\"os parameter from our torque measurements via:
\begin{align}
\eta=\frac{\tau_{in}}{m\ r\ a_g}
\end{align}
where $a_g$ is the gravitational acceleration towards the sun. Our results yield a 95\%-confidence limit on the E\"otv\"os parameter for superconducting niobium versus copper of 
\begin{equation}
    \eta_{\text{Nb*-Cu}}~\leq~2.0~\times~10^{-9}.
\end{equation}

It has been suggested that the material of interest is not the bulk superconductor but only the Cooper pairs \cite{superTheory}. Additionally, parameterizing in terms of Cooper pair number allows for comparison between different superconducting materials. 

The total mass of Cooper pairs within a test body is:
\begin{align}
    m_{CP}\approx\frac{m_e}{m_n} \frac{N_e}{N_n} \frac{N_v}{N_e} \frac{N_{CP}}{N_v}m \label{mCP}
\end{align}
where $m_e$ and $m_p$ are respectively the mass of the electron and proton, $N_e$, $N_n$, $N_v$, and $N_{CP}$ are the number of electrons, nucleons, valence electrons, and electrons in Cooper pairs, respectively. 
The ratio $N_{CP}/N_v$ was estimated with the two fluid model \cite{sharma2021superconductivity}:
\begin{equation}
    \frac{N_{CP}}{N_v}=1-\bigg( \frac{T}{T_c}\bigg)^4
\end{equation}
where $T$ is the temperature and $T_c$ is the critical temperature. We use a conservative upper limit of the temperature of the pendulum of 8 K which corresponds to a Cooper pair fraction of $N_{CP}/N_v=0.38$.



Using Equation \ref{mCP}, we convert the measurement for superconducting niobium to a 95\%-confidence limit on the E\"otv\"os parameter for Cooper pairs versus unpaired electrons of:
\begin{equation}
    \eta_{\text{CP-ee}}\leq 9.2\times 10^{-4}.
\end{equation}

\textit{Conclusion} - We used a cryogenic torsion pendulum with superconducting-niobium and copper test bodies to constrain deviations from the WEP for superconducting niobium and Cooper pairs with 95\%-confidence to $\eta_{\text{Nb*-Cu}}~\leq~2.0\times10^{-9}$ and $\eta_{\text{CP-ee}}\leq9.2\times10^{-4}$, respectively. This improves upon the result reported by \citet{Tajmar} by more than two orders of magnitude. Our experiment places strong constraints on any enhanced gravitational effects of superconductors.

\textit{Data Availability} - Code to generate the results shown here can be found at \url{https://github.com/EotWash/SuperEP}. Data avaible upon request.

\textit{Acknowledgements} - Special thanks to Andreas Schilling for help with superconductor physics. Participation from the University of Washington, Seattle, was supported by funding from the NSF under Awards PHY-1607385, PHY-1607391, PHY-1912380, and PHY-1912514.

\bibliographystyle{unsrtnat}
\bibliography{SuperconEP.bib}

\begin{thebibliography}{16}
\providecommand{\natexlab}[1]{#1}
\providecommand{\url}[1]{\texttt{#1}}
\expandafter\ifx\csname urlstyle\endcsname\relax
  \providecommand{\doi}[1]{doi: #1}\else
  \providecommand{\doi}{doi: \begingroup \urlstyle{rm}\Url}\fi

\bibitem[v.~E\"otv\"os et~al.(1922)v.~E\"otv\"os, Pek\'ar, and Fekete]{Eotvos}
Roland v.~E\"otv\"os, Desiderius Pek\'ar, and Eugen Fekete.
\newblock Beitr\"age zum gesetze der proportionalit\"at von tr\"agheit und gravit\"at.
\newblock \emph{Annalen der Physik}, 373\penalty0 (9):\penalty0 11--66, 1922.
\newblock \doi{10.1002/andp.19223730903}.
\newblock URL \url{https://onlinelibrary.wiley.com/doi/abs/10.1002/andp.19223730903}.

\bibitem[Wagner et~al.(2012)Wagner, Schlamminger, Gundlach, and Adelberger]{NewWash}
T~A Wagner, S~Schlamminger, J~H Gundlach, and E~G Adelberger.
\newblock Torsion-balance tests of the weak equivalence principle.
\newblock \emph{Classical and Quantum Gravity}, 29\penalty0 (18):\penalty0 184002, Aug 2012.
\newblock \doi{10.1088/0264-9381/29/18/184002}.
\newblock URL \url{https://dx.doi.org/10.1088/0264-9381/29/18/184002}.

\bibitem[Williams et~al.(2012)Williams, Turyshev, and Boggs]{LLR}
James~G Williams, Slava~G Turyshev, and Dale~H Boggs.
\newblock Lunar laser ranging tests of the equivalence principle.
\newblock \emph{Classical and Quantum Gravity}, 29\penalty0 (18):\penalty0 184004, Aug 2012.
\newblock \doi{10.1088/0264-9381/29/18/184004}.
\newblock URL \url{https://doi.org/10.1088\%2F0264-9381\%2F29\%2F18\%2F184004}.

\bibitem[Touboul et~al.(2022)Touboul, M\'etris, Rodrigues, Berg\'e, Robert, Baghi, Andr\'e, Bedouet, Boulanger, Bremer, Carle, Chhun, Christophe, Cipolla, Damour, Danto, Demange, Dittus, Dhuicque, Fayet, Foulon, Guidotti, Hagedorn, Hardy, Huynh, Kayser, Lala, L\"ammerzahl, Lebat, Liorzou, List, L\"offler, Panet, Pernot-Borr\`as, Perraud, Pires, Pouilloux, Prieur, Rebray, Reynaud, Rievers, Selig, Serron, Sumner, Tanguy, Torresi, and Visser]{MICROSCOPE}
Pierre Touboul, Gilles M\'etris, Manuel Rodrigues, Joel Berg\'e, Alain Robert, Quentin Baghi, Yves Andr\'e, Judica\"el Bedouet, Damien Boulanger, Stefanie Bremer, Patrice Carle, Ratana Chhun, Bruno Christophe, Valerio Cipolla, Thibault Damour, Pascale Danto, Louis Demange, Hansjoerg Dittus, Oc\'eane Dhuicque, Pierre Fayet, Bernard Foulon, Pierre-Yves Guidotti, Daniel Hagedorn, Emilie Hardy, Phuong-Anh Huynh, Patrick Kayser, St\'ephanie Lala, Claus L\"ammerzahl, Vincent Lebat, Fran\ifmmode \mbox{\c{c}}\else~\c{c}\fi{}oise Liorzou, Meike List, Frank L\"offler, Isabelle Panet, Martin Pernot-Borr\`as, Laurent Perraud, Sandrine Pires, Benjamin Pouilloux, Pascal Prieur, Alexandre Rebray, Serge Reynaud, Benny Rievers, Hanns Selig, Laura Serron, Timothy Sumner, Nicolas Tanguy, Patrizia Torresi, and Pieter Visser.
\newblock {MICROSCOPE} mission: Final results of the test of the equivalence principle.
\newblock \emph{Phys. Rev. Lett.}, 129:\penalty0 121102, Sep 2022.
\newblock \doi{10.1103/PhysRevLett.129.121102}.
\newblock URL \url{https://link.aps.org/doi/10.1103/PhysRevLett.129.121102}.

\bibitem[de~Matos(2010)]{superTheory}
Clovis~Jacinto de~Matos.
\newblock Physical vacuum in superconductors.
\newblock \emph{Journal of Superconductivity and Novel Magnetism}, 23\penalty0 (8):\penalty0 1443--1453, Dec 2010.
\newblock ISSN 1557-1947.
\newblock \doi{10.1007/s10948-010-0793-x}.
\newblock URL \url{https://doi.org/10.1007/s10948-010-0793-x}.

\bibitem[Quach(2015)]{GCasimir}
James~Q. Quach.
\newblock Gravitational casimir effect.
\newblock \emph{Phys. Rev. Lett.}, 114:\penalty0 081104, Feb 2015.
\newblock \doi{10.1103/PhysRevLett.114.081104}.
\newblock URL \url{https://link.aps.org/doi/10.1103/PhysRevLett.114.081104}.

\bibitem[Minter et~al.(2010)Minter, Wegter-McNelly, and Chiao]{GWReflect}
Stephen~J. Minter, Kirk Wegter-McNelly, and Raymond~Y. Chiao.
\newblock Do mirrors for gravitational waves exist?
\newblock \emph{Physica E: Low-dimensional Systems and Nanostructures}, 42\penalty0 (3):\penalty0 234 -- 255, 2010.
\newblock ISSN 1386-9477.
\newblock \doi{https://doi.org/10.1016/j.physe.2009.06.056}.
\newblock URL \url{http://www.sciencedirect.com/science/article/pii/S1386947709001994}.
\newblock Proceedings of the international conference Frontiers of Quantum and Mesoscopic Thermodynamics FQMT '08.

\bibitem[Jain et~al.(1987)Jain, Lukens, and Tsai]{Jain}
A.~K. Jain, J.~E. Lukens, and J.~S. Tsai.
\newblock Test for relativistic gravitational effects on charged particles.
\newblock \emph{Phys. Rev. Lett.}, 58:\penalty0 1165--1168, Mar 1987.
\newblock \doi{10.1103/PhysRevLett.58.1165}.
\newblock URL \url{https://link.aps.org/doi/10.1103/PhysRevLett.58.1165}.

\bibitem[Tajmar et~al.(2009)Tajmar, Plesescu, and Seifert]{Tajmar}
M~Tajmar, F~Plesescu, and B~Seifert.
\newblock Measuring the dependence of weight on temperature in the low-temperature regime using a magnetic suspension balance.
\newblock \emph{Measurement Science and Technology}, 21\penalty0 (1):\penalty0 015111, dec 2009.
\newblock \doi{10.1088/0957-0233/21/1/015111}.
\newblock URL \url{https://doi.org/10.1088\%2F0957-0233\%2F21\%2F1\%2F015111}.

\bibitem[Newman et~al.(20014)Newman, Bantel, Berg, and Cross]{otherCryo}
Riley Newman, Michael Bantel, Eric Berg, and William Cross.
\newblock A measurement of {G} with a cryogenic torsion pendulum.
\newblock \emph{Philosophical Transactions of the Royal Society A: Mathematical, Physical and Engineering Sciences}, 372, 20014.
\newblock \doi{10.1088/0957-0233/21/1/015111}.
\newblock URL \url{http://doi.org/10.1098/rsta.2014.0025}.

\bibitem[Fleischer et~al.(2022)Fleischer, Ross, Venkateswara, Hagedorn, Shaw, Swanson, Heckel, and Gundlach]{RSI}
S.~M. Fleischer, M.~P. Ross, K.~Venkateswara, C.~A. Hagedorn, E.~A. Shaw, E.~Swanson, B.~R. Heckel, and J.~H. Gundlach.
\newblock A cryogenic torsion balance using a liquid-cryogen free, ultra-low vibration cryostat.
\newblock \emph{Review of Scientific Instruments}, 93\penalty0 (6):\penalty0 064505, 2022.
\newblock \doi{10.1063/5.0089933}.
\newblock URL \url{https://doi.org/10.1063/5.0089933}.

\bibitem[Matthias et~al.(1963)Matthias, Geballe, and Compton]{Nb}
B.~T. Matthias, T.~H. Geballe, and V.~B. Compton.
\newblock Superconductivity.
\newblock \emph{Rev. Mod. Phys.}, 35:\penalty0 1--22, Jan 1963.
\newblock \doi{10.1103/RevModPhys.35.1}.
\newblock URL \url{https://link.aps.org/doi/10.1103/RevModPhys.35.1}.

\bibitem[Arp et~al.(2013)Arp, Hagedorn, Schlamminger, and Gundlach]{MSA}
T.~B. Arp, C.~A. Hagedorn, S.~Schlamminger, and J.~H. Gundlach.
\newblock A reference-beam autocollimator with nanoradian sensitivity from mhz to khz and dynamic range of 107.
\newblock \emph{Review of Scientific Instruments}, 84\penalty0 (9):\penalty0 095007, 2013.
\newblock \doi{10.1063/1.4821653}.
\newblock URL \url{https://doi.org/10.1063/1.4821653}.

\bibitem[{Astropy Collaboration} et~al.(2013){Astropy Collaboration}, {Robitaille}, {Tollerud}, {Greenfield}, {Droettboom}, {Bray}, {Aldcroft}, {Davis}, {Ginsburg}, {Price-Whelan}, {Kerzendorf}, {Conley}, {Crighton}, {Barbary}, {Muna}, {Ferguson}, {Grollier}, {Parikh}, {Nair}, {Unther}, {Deil}, {Woillez}, {Conseil}, {Kramer}, {Turner}, {Singer}, {Fox}, {Weaver}, {Zabalza}, {Edwards}, {Azalee Bostroem}, {Burke}, {Casey}, {Crawford}, {Dencheva}, {Ely}, {Jenness}, {Labrie}, {Lim}, {Pierfederici}, {Pontzen}, {Ptak}, {Refsdal}, {Servillat}, and {Streicher}]{astropy1}
{Astropy Collaboration}, T.~P. {Robitaille}, E.~J. {Tollerud}, P.~{Greenfield}, M.~{Droettboom}, E.~{Bray}, T.~{Aldcroft}, M.~{Davis}, A.~{Ginsburg}, A.~M. {Price-Whelan}, W.~E. {Kerzendorf}, A.~{Conley}, N.~{Crighton}, K.~{Barbary}, D.~{Muna}, H.~{Ferguson}, F.~{Grollier}, M.~M. {Parikh}, P.~H. {Nair}, H.~M. {Unther}, C.~{Deil}, J.~{Woillez}, S.~{Conseil}, R.~{Kramer}, J.~E.~H. {Turner}, L.~{Singer}, R.~{Fox}, B.~A. {Weaver}, V.~{Zabalza}, Z.~I. {Edwards}, K.~{Azalee Bostroem}, D.~J. {Burke}, A.~R. {Casey}, S.~M. {Crawford}, N.~{Dencheva}, J.~{Ely}, T.~{Jenness}, K.~{Labrie}, P.~L. {Lim}, F.~{Pierfederici}, A.~{Pontzen}, A.~{Ptak}, B.~{Refsdal}, M.~{Servillat}, and O.~{Streicher}.
\newblock {Astropy: A community Python package for astronomy}.
\newblock \emph{Astronomy \& Astrophysics}, 558:\penalty0 A33, October 2013.
\newblock \doi{10.1051/0004-6361/201322068}.

\bibitem[{Astropy Collaboration} et~al.(2018){Astropy Collaboration}, {Price-Whelan}, {Sip{\H{o}}cz}, {G{\"u}nther}, {Lim}, {Crawford}, {Conseil}, {Shupe}, {Craig}, {Dencheva}, {Ginsburg}, {Vand erPlas}, {Bradley}, {P{\'e}rez-Su{\'a}rez}, {de Val-Borro}, {Aldcroft}, {Cruz}, {Robitaille}, {Tollerud}, {Ardelean}, {Babej}, {Bach}, {Bachetti}, {Bakanov}, {Bamford}, {Barentsen}, {Barmby}, {Baumbach}, {Berry}, {Biscani}, {Boquien}, {Bostroem}, {Bouma}, {Brammer}, {Bray}, {Breytenbach}, {Buddelmeijer}, {Burke}, {Calderone}, {Cano Rodr{\'\i}guez}, {Cara}, {Cardoso}, {Cheedella}, {Copin}, {Corrales}, {Crichton}, {D'Avella}, {Deil}, {Depagne}, {Dietrich}, {Donath}, {Droettboom}, {Earl}, {Erben}, {Fabbro}, {Ferreira}, {Finethy}, {Fox}, {Garrison}, {Gibbons}, {Goldstein}, {Gommers}, {Greco}, {Greenfield}, {Groener}, {Grollier}, {Hagen}, {Hirst}, {Homeier}, {Horton}, {Hosseinzadeh}, {Hu}, {Hunkeler}, {Ivezi{\'c}}, {Jain}, {Jenness}, {Kanarek}, {Kendrew}, {Kern}, {Kerzendorf}, {Khvalko}, {King}, {Kirkby}, {Kulkarni},
  {Kumar}, {Lee}, {Lenz}, {Littlefair}, {Ma}, {Macleod}, {Mastropietro}, {McCully}, {Montagnac}, {Morris}, {Mueller}, {Mumford}, {Muna}, {Murphy}, {Nelson}, {Nguyen}, {Ninan}, {N{\"o}the}, {Ogaz}, {Oh}, {Parejko}, {Parley}, {Pascual}, {Patil}, {Patil}, {Plunkett}, {Prochaska}, {Rastogi}, {Reddy Janga}, {Sabater}, {Sakurikar}, {Seifert}, {Sherbert}, {Sherwood-Taylor}, {Shih}, {Sick}, {Silbiger}, {Singanamalla}, {Singer}, {Sladen}, {Sooley}, {Sornarajah}, {Streicher}, {Teuben}, {Thomas}, {Tremblay}, {Turner}, {Terr{\'o}n}, {van Kerkwijk}, {de la Vega}, {Watkins}, {Weaver}, {Whitmore}, {Woillez}, {Zabalza}, and {Astropy Contributors}]{astropy2}
{Astropy Collaboration}, A.~M. {Price-Whelan}, B.~M. {Sip{\H{o}}cz}, H.~M. {G{\"u}nther}, P.~L. {Lim}, S.~M. {Crawford}, S.~{Conseil}, D.~L. {Shupe}, M.~W. {Craig}, N.~{Dencheva}, A.~{Ginsburg}, J.~T. {Vand erPlas}, L.~D. {Bradley}, D.~{P{\'e}rez-Su{\'a}rez}, M.~{de Val-Borro}, T.~L. {Aldcroft}, K.~L. {Cruz}, T.~P. {Robitaille}, E.~J. {Tollerud}, C.~{Ardelean}, T.~{Babej}, Y.~P. {Bach}, M.~{Bachetti}, A.~V. {Bakanov}, S.~P. {Bamford}, G.~{Barentsen}, P.~{Barmby}, A.~{Baumbach}, K.~L. {Berry}, F.~{Biscani}, M.~{Boquien}, K.~A. {Bostroem}, L.~G. {Bouma}, G.~B. {Brammer}, E.~M. {Bray}, H.~{Breytenbach}, H.~{Buddelmeijer}, D.~J. {Burke}, G.~{Calderone}, J.~L. {Cano Rodr{\'\i}guez}, M.~{Cara}, J.~V.~M. {Cardoso}, S.~{Cheedella}, Y.~{Copin}, L.~{Corrales}, D.~{Crichton}, D.~{D'Avella}, C.~{Deil}, {\'E}.~{Depagne}, J.~P. {Dietrich}, A.~{Donath}, M.~{Droettboom}, N.~{Earl}, T.~{Erben}, S.~{Fabbro}, L.~A. {Ferreira}, T.~{Finethy}, R.~T. {Fox}, L.~H. {Garrison}, S.~L.~J. {Gibbons}, D.~A. {Goldstein}, R.~{Gommers},
  J.~P. {Greco}, P.~{Greenfield}, A.~M. {Groener}, F.~{Grollier}, A.~{Hagen}, P.~{Hirst}, D.~{Homeier}, A.~J. {Horton}, G.~{Hosseinzadeh}, L.~{Hu}, J.~S. {Hunkeler}, {\v{Z}}.~{Ivezi{\'c}}, A.~{Jain}, T.~{Jenness}, G.~{Kanarek}, S.~{Kendrew}, N.~S. {Kern}, W.~E. {Kerzendorf}, A.~{Khvalko}, J.~{King}, D.~{Kirkby}, A.~M. {Kulkarni}, A.~{Kumar}, A.~{Lee}, D.~{Lenz}, S.~P. {Littlefair}, Z.~{Ma}, D.~M. {Macleod}, M.~{Mastropietro}, C.~{McCully}, S.~{Montagnac}, B.~M. {Morris}, M.~{Mueller}, S.~J. {Mumford}, D.~{Muna}, N.~A. {Murphy}, S.~{Nelson}, G.~H. {Nguyen}, J.~P. {Ninan}, M.~{N{\"o}the}, S.~{Ogaz}, S.~{Oh}, J.~K. {Parejko}, N.~{Parley}, S.~{Pascual}, R.~{Patil}, A.~A. {Patil}, A.~L. {Plunkett}, J.~X. {Prochaska}, T.~{Rastogi}, V.~{Reddy Janga}, J.~{Sabater}, P.~{Sakurikar}, M.~{Seifert}, L.~E. {Sherbert}, H.~{Sherwood-Taylor}, A.~Y. {Shih}, J.~{Sick}, M.~T. {Silbiger}, S.~{Singanamalla}, L.~P. {Singer}, P.~H. {Sladen}, K.~A. {Sooley}, S.~{Sornarajah}, O.~{Streicher}, P.~{Teuben}, S.~W. {Thomas}, G.~R.
  {Tremblay}, J.~E.~H. {Turner}, V.~{Terr{\'o}n}, M.~H. {van Kerkwijk}, A.~{de la Vega}, L.~L. {Watkins}, B.~A. {Weaver}, J.~B. {Whitmore}, J.~{Woillez}, V.~{Zabalza}, and {Astropy Contributors}.
\newblock {The Astropy Project: Building an Open-science Project and Status of the v2.0 Core Package}.
\newblock \emph{The Astronomical Journal}, 156\penalty0 (3):\penalty0 123, September 2018.
\newblock \doi{10.3847/1538-3881/aabc4f}.

\bibitem[Sharma(2021)]{sharma2021superconductivity}
Ram~Gopal Sharma.
\newblock \emph{Superconductivity: Basics and applications to magnets}, volume 214.
\newblock Springer Nature, 2021.

\end{thebibliography}

\end{document}